\documentclass[preprint,prd]{revtex4}
\usepackage{graphicx}

\begin{document}

\newcommand{\be}{\begin{equation}}
\newcommand{\ee}{\end{equation}}
\newcommand{\ben}{\begin{eqnarray}}
\newcommand{\een}{\end{eqnarray}}

\newcommand{\la}{{\lambda}}
\newcommand{\Om}{{\Omega}}
\newcommand{\ta}{{\tilde a}}
\newcommand{\bg}{{\bar g}}
\newcommand{\bh}{{\bar h}}
\newcommand{\si}{{\sigma}}
\newcommand{\C}{{\cal C}}
\newcommand{\D}{{\cal D}}
\newcommand{\cA}{{\cal A}}
\newcommand{\cT}{{\cal T}}
\newcommand{\cO}{{\cal O}}
\newcommand{\eeo}{\cO ({1 \over E})}
\newcommand{\cR}{{\cal R}}
\newcommand{\cL}{{\cal L}}
\newcommand{\T}{{\cal T}}
\newcommand{\M}{{\cal M}}

\newcommand{\p}{\partial}
\newcommand{\na}{\nabla}
\newcommand{\ssum}{\sum\limits_{i = 1}^3}
\newcommand{\dssum}{\sum\limits_{i = 1}^2}
\newcommand{\tal}{{\tilde \alpha}}

\newcommand{\tp}{{\tilde \phi}}
\newcommand{\tPhi}{\tilde \Phi}
\newcommand{\tpsi}{\tilde \psi}
\newcommand{\tim}{{\tilde \mu}}
\newcommand{\tr}{{\tilde \rho}}
\newcommand{\tir}{{\tilde r}}
\newcommand{\rp}{r_{+}}
\newcommand{\hr}{{\hat r}}
\newcommand{\rv}{{r_{v}}}
\newcommand{\dr}{\frac{d}{d \hr}}
\newcommand{\dR}{\frac{d}{d R}}

\newcommand{\hhf}{{\hat \phi}}
\newcommand{\hhM}{{\hat M}}
\newcommand{\hhQ}{{\hat Q}}
\newcommand{\hht}{{\hat t}}
\newcommand{\hhr}{{\hat r}}
\newcommand{\hhS}{{\hat \Sigma}}
\newcommand{\hhD}{{\hat \Delta}}
\newcommand{\hhm}{{\hat \mu}}
\newcommand{\hro}{{\hat \rho}}
\newcommand{\hhz}{{\hat z}}

\newcommand{\tD}{{\tilde D}}
\newcommand{\tB}{{\tilde B}}
\newcommand{\tV}{{\tilde V}}
\newcommand{\hT}{\hat T}
\newcommand{\tE}{\tilde E}
\newcommand{\tT}{\tilde T}
\newcommand{\hC}{\hat C}
\newcommand{\ep}{\epsilon}
\newcommand{\bep}{\bar \epsilon}
\newcommand{\ppp}{\varphi}
\newcommand{\Ga}{\Gamma}
\newcommand{\ga}{\gamma}
\newcommand{\hth}{\hat \theta}

\title{Thick Domain Walls in AdS Black Hole Spacetimes}
\date{\today}

\author{Rafa{\l} Moderski}
\affiliation{Nicolaus Copernicus Astronomical Center \protect \\
Polish Academy of Sciences \protect \\
00-716 Warsaw, Bartycka 18, Poland \protect \\
moderski@camk.edu.pl}

\author{Marek Rogatko}
\affiliation{Institute of Physics \protect \\
Maria Curie-Sklodowska University \protect \\
20-031 Lublin, pl.~Marii Curie-Sklodowskiej 1, Poland \protect \\
rogat@kft.umcs.lublin.pl}

\begin{abstract}
Equations of motion for a real self-gravitating scalar field in the
background of a black hole with negative cosmological constant were
solved numerically.  We obtain a sequence of static axisymmetric
solutions representing thick domain wall cosmological black hole
systems, depending on the mass of black hole, cosmological parameter
and the parameter binding black hole mass with the width of the domain
wall.  For the case of extremal cosmological black hole the expulsion
of scalar field from the black hole strongly depends on it.
\end{abstract}

\maketitle

\section{Introduction}
The early Universe and the phase transitions there provide signs of
the high-energy phenomena which are beyond the range of contemporary
accelerators ~\cite{vs94}.  Such topological defects as cosmic strings
attracted a great interests and were widely studied in literature.
Black hole cosmic string configurations studies revealed the evidences
that cosmic string could thread it or could be expelled from black
hole.  In the case of extreme Reissner-Nordstr\"om (RN) black holes it
was found that there was a range of the black hole parameters for
which the expulsion of the vortex took
place~\cite{dgt92,agk95,cha98a,cha98b,bg98,beg99}.  In dilaton gravity
theory being the low-energy limit of the string theory it was also
justified~\cite{mr98a,mr98b,mr99,sg00} that a vortex could be treated
by the remote observer as a hair on the black hole.  Moreover, extreme
dilaton black hole always expel the Higgs field, one has to do with
the so-called \textit{Meissner effect}.  The problem of vortices in de
Sitter background was analyzed in Ref.~\cite{gm02a}, while the
behaviour of Abelian Higgs vortex solutions in Schwarzschild anti de
Sitter (AdS), Kerr, Kerr-AdS and Reissner-Nordstr\"om AdS (RN-AdS) was
studied in Refs.~\cite{dgm02,gm02b}.  The vortex solution for Abelian
Higgs field equations in the background of four-dimensional black
string was considered in Ref.~\cite{dj02}.

The idea that the Universe is embedded in higher-dimensional spacetime
acquires much attention.  The resurgence is motivated by the
possibility of resolving the hierarchy problem~\cite{rs99a,rs99b},
i.e., the difference in magnitudes of the Planck scale and the
electroweak scales.  Also the most promising candidate for a unified
theory of Nature, superstring theory predicts the existence of the
so-called D-branes which renew in turn the idea of \textit{brane
worlds}.  From this theory point of view the model of brane world
stems from the idea presented by Horava and Witten~\cite{hw96a,hw96b}.
Namely, the strong coupling limit of $E_{8} \times E_{8}$ heterotic
string theory at low energy is described by eleven-dimensional
supergravity with eleventh dimension compactified on an orbifold with
$Z_{2}$ symmetry.  The two boundaries of spacetime are ten-dimensional
planes, to which gauge theories are confined.
Next~\cite{wit96,luk99,low99} it was argued that six of the eleven
dimensions could be consistently compactified and in the limit
spacetime looked five-dimensional with four-dimensional boundary
brane.

The studies of the interplay between black holes and domain walls
(branes) acquire more attention.  The problem of stability of a
Nambu-Goto membrane in RN-dS spacetime was studied in~\cite{hii01}.
The gravitationally interacting system of a thick domain wall and
Schwarschild black hole was considered in~\cite{mor00,mor03}.  Emparan
et al.~\cite{egs01} elaborated the problem of a black hole on a
topological domain wall.

In~\cite{rog01} the dilaton black hole-domain wall system was studied
analytically and it was revealed that for the extremal dilaton black
hole one had to do with the expulsion of the scalar field from the
black hole (the so-called \textit{Meissner effect}).  The numerical
studies of domain wall in the spacetime of dilaton black hole was
considered in Ref.~\cite{mr03}, where the thickness of the domain wall
and a potential of the scalar field $\varphi^4$ and sine-Gordon were
taken into account.  In the case of a real self-interacting scalar
field in the background of RN solution it was found that in the
extremal case there was a parameter depending on black hole mass and
the width of the domain wall which constituted the upper limit for the
expulsion to occur~\cite{mr04}.  Dynamics of domain walls intersecting
black holes, taking into account the evolution of the system during
the separation was elaborated in~\cite{fla06}.

The importance of introducing the cosmological into considerations
comes not only from the theoretical point of view but also for the
observational results of our Universe.  Small cosmological constant
may be an alternative to the dark energy in explanation the observed a
type Ia suprenova acceleration.  On the other hand, as far as a
negative cosmological constant is concerned a tremendous interest has
focused on issues related to the AdS spacetime.  One of them is the
AdS/CFT (conformal field theory) correspondence which states that
conformal field theories in d-dimension $R_{d}$ are described in terms
of supergravity or string theory on the product spacetime consisting
with asymptotically $AdS_{d+1}$ and compact manifold, providing that
there are relations between data on the boundary $R_{d}$ of the
$AdS_{d+1}$ and the data in the bulk $AdS_{d+1}$
\cite{mal98,wit98a,wit98b}.

Our paper will be devoted to the behaviour of the AdS black hole thick
domain wall system.  A domain wall will be simulated by a
self-interacting scalar field.  We shall numerically analyze the
behaviour of scalar fields for various value of cosmological
parameter.  The brief outline of our paper is the following. The next
section is devoted to the basic equations of the considered problem.
In Sec.~\ref{sec:boundry}-\ref{sec:numeric} we presented the boundary
conditions of the problem as well as the numerical analysis of the
equations of motion for the two cases of potentials with discrete sets
of minima.  Namely, we shall analyze the $\varphi^4$ and sine-Gordon
potentials.  The conclusions and discussions will be presented in
Sec.~\ref{sec:conc}.

\section{The basic equations of the problem}
In our paper we shall consider a spherically symmetric static black
hole with negative cosmological constant.  The metric of which is
written as follows:
\be ds^2 = - \left ( 1 - {2 M \over r} + {Q \over r^2} -{ \Lambda
\over 3} r^2 \right ) dt^2 + {d r^2 \over {\left ( 1 - {2 M \over r} +
{Q \over r^2} -{ \Lambda \over 3} r^2 \right )}} + r^2 (d \theta^2 +
\sin^2 \theta d \ppp^2), \ee
where we define $\Lambda = { - 3 \over l^2}$ and $Q$ is the charge of
the considered black hole.  In what follows we shall denote by $V(r) =
1 - {2 M \over r} + {Q \over r^2} -{ \Lambda \over 3} r^2$.  The
condition $V(r) = 0$ is a quadratic algebraic equation for $r$.  As
was mentioned in~\cite{rom92} the standard closed-form for the above
mentioned four roots is rather lengthy and not especially
illuminating.  In what follows the AdS black hole spacetime will be
the background metric on which one solves the domain wall Eqs. of
motion.  The domain wall will be simulated by a self-interacting
scalar field.

A general matter Lagrangian with real Higgs field and the symmetry
breaking potential of the form will be taken into account
\be {\cal L}_{dw} = - {1 \over 2} \na_{\mu} \varphi \na^{\mu} \varphi
- U(\varphi), \ee
where by $\varphi$ we have denoted the real Higgs field, while by
$U(\varphi)$ the symmetry breaking potential having a discrete set of
degenerate minima.  The energy-momentum tensor for the domain wall
yields
\be T_{ij}(\varphi) = - {1 \over 2} g_{ij} \na_{m} \varphi \na^{m}
\varphi - U(\varphi) g_{ij} + \na_{i} \varphi \na_{j} \varphi.
\label{ten}
\ee
For the convenience we scale out parameters via transformation $ X =
{\varphi / \eta}$ and $\ep = 8 \pi G \eta^2$.  The parameter $\ep$
represents the gravitational strength and is connected with the
gravitational interaction of the Higgs field.  Defining $V(X) =
{U(\varphi) \over V_{F}}$, where $V_{F} = \lambda \eta^4$ we arrive at
the following expression:
\be 8 \pi G {\cal L}_{dw} = - {\ep \over w^2} \bigg[ w^2 {\na_{\mu} X
\na^{\mu} X \over 2} + V(X) \bigg],
\label{dw}
\ee
where $w = \sqrt{{\ep \over 8 \pi G V_{F}}}$ represents the inverse
mass of the scalar after symmetry breaking, which also characterize
the width of the wall defect within the theory under consideration.
Having in mind (\ref{dw}) the equations for $X$ field may be written
as follows:
\be \na_{\mu} \na^{\mu} X - {1 \over w^2}{ \p V \over \p X} = 0.\ee
In our considerations we take into account for two cases of potentials
with a discrete set of degenerate minima, namely

\noindent
the $\varphi^4$ potential described by the following equation:
\be U_1(\varphi) = {\lambda \over 4} (\varphi^2 - \eta^2)^2,
\label{phi4}
\ee
and the sine-Gordon potential of the form
\be U_2(\varphi) = \lambda \eta^4 \left [ 1 + \cos(\varphi/\eta)
\right ].
\label{sinG}
\ee

\section{\label{sec:boundry}The boundary conditions}
In the background of the black hole spacetime the equation of motion
for the scalar field $X$ implies
\be {1 \over r^2} \p_{r} \bigg[ \big ( r^2 - 2 Mr + Q^2 - {\Lambda
\over 3}r^4 \big) \p_{r} X \bigg] + {1 \over r^2 \sin \theta }
\p_{\theta} \bigg[ \sin \theta \p_{\theta} X \bigg] = {1 \over w^2} {
\p V \over \p X}.
\label{motion}
\ee
Consequently, having in mind relation (\ref{ten}) one can define the
following quantity for the scalar field $\varphi$:
\be \tE = {T_{t}{}{}^{t} \over \la \eta^4} = \bigg[ - {1 \over 2}
\big( X_{,r} \big)^2 \bigg( 1 - {2M \over r} + {Q \over r^2} -
{\Lambda \over 3} r^2 \bigg) -{ 1 \over 2} \big( X_{,\theta} \big)^2
{1 \over r^2} \bigg] { w^2} - V(X).
\label{energy}
\ee
As we have to do with non-asymptotically flat spacetime it cannot be
interpreted as an energy density of scalar field.  On the horizon of
the black hole one has the following boundary conditions:
\be V'(r_{BH}) \p_{r} X \mid_{r = r_{BH}} = - {1 \over r_{BH}^2 \sin
\theta} \p_{\theta} \bigg[ \sin \theta \p_{\theta} X \bigg] + {1 \over
w^2} {\p V \over \p X}.  \ee
As in Refs.~\cite{mr03,mr04} we shall restrict our investigations to
the case when the core of the domain wall is located in the equatorial
plane, i.e., $\theta = \pi / 2$. One ought also to impose the
Dirichlet boundary conditions at the equatorial plane as follows:
\be X \mid_{\theta = {\pi \over 2}} = 0, \ee
and the regularity condition for scalar field on the symmetry axis
requiring the Neumann boundary condition on $z$-axis in the form as
\be {\p X \over \p \theta} \mid_{\theta = 0} = 0.  \ee
In order to find the solution of the problem our first task will be to
establish the asymptotically behaviour of the scalar field at the
large distances. Further it will be necessary to find the numerical
solution of its equation of motion.

In this case at large distances from the horizon we demand that the
black hole domain wall solution ought to be solution of the domain
wall in AdS or dS spacetime.  Defining the metric as follows:
\be ds^2 = - \bigg( 1 + {k r^2 \over l^2} \bigg) dt^2 + {dr^2 \over
\bigg( 1 + {k r^2 \over l^2} \bigg)} + r^2 (d \theta^2 + \sin^2 \theta
d \ppp^2), \ee
where $k = - 1$ we have for AdS spacetime, on the other hand $k = 1$
is responsible for dS one, the equation of motion for the scalar field
$X$ yields
\be \bigg( 1 + {k r^2 \over l^2} \bigg) \p_{r}^{2} X + {2 \over r}
\bigg( 1 + {2 kr^2 \over l^2} \bigg) \p_{r} X + {1 \over r^2}
\p_{\theta}^{2} X + {1 \over r^2} \cot \theta~ \p_{\theta} X - {1
\over w^2} {\p V \over \p X} = 0.  \ee
As in Ref.~\cite{gm02a} we set $\rho = r \cos \theta$ and the scalar
field $X$ is a function of $\rho$.  By virtue of this equations of
motion may be written in the form as
\be \bigg( 1 + {k \rho^2 \over l^2} \bigg) X_{, \rho \rho} + {4 k \rho
\over l^2} X_{,\rho} - {1 \over w^2} {\p V \over \p X} = 0.
\label{xxx}
\ee
Let us solve the equation for the potential $V(X) = {1 \over 4} (X^2 -
1)^2$.  The analitycal solution of Eq.~(\ref{xxx}) is not known, but
we shall search for the solution for magnitude of scalar field
$X(\rho)$ at constant large distances, i.e., for $\rho \rightarrow
\infty$.  It follows in particular that $X( \rho \rightarrow \infty )
= 1 = X_{0}$. Equation~(\ref{xxx}) is approximately satisfied by
$X_{0} \simeq 1 $, where $X_{0}$ is the minimum of the potential.
Denoting fluctuations about this minimum by $\psi (\rho) $ we have the
following:
\be X(\rho) = X_{0} + \psi (\rho).  \ee
It can be verified by expanding the scalar field $X$ in the equation
of motion that it reduces to the form
\be k \rho^2 \psi_{, \rho \rho} + 4 k \rho \psi_{, \rho} + {l^2 \over
w^2} \psi = 0.
\label{skr}
\ee
In deriving Eq.~(\ref{skr}) one has neglected terms of order of unity
in coefficients in first and second terms with respect to the terms
involving derivatives of scalar field connected with ${\rho^2 \over
l^2}$.

Thus, Eq.~(\ref{skr}) for large $\rho \rightarrow \infty$ has the
approximate solution in the form of
\be X(\rho) \simeq X_{0} \bigg[ 1 - \bigg( {\rho \over \rho_{0}}
\bigg)^{{- 3 + \sqrt{9 - 4 \beta} \over 2}} \bigg], \ee
where $\beta = {l^2 /w^2 k}$, in our case $\beta < 0$.  The same
procedure applied to the sine-Gordon potential $V(X) = 1 + \cos X$
case reveals the following relation:
\be X(\rho) \simeq X_{0} \bigg[ 1 - \bigg( {\rho \over \rho_{0}}
\bigg)^{- 2 + \sqrt{4 - \beta}} \bigg].  \ee

\section{\label{sec:numeric}Numerical calculations}

\subsection{Cosmological AdS black hole -- $\phi^4$ potential}
Having laid out for our considerations, we proceed to describe the
numerical analysis of the problem.  First, we convert to equation of
motion (\ref{motion}) to the form more suitable for finite
differencing, namely one has the following:
\be \partial_r \left[ \left( r^2 - 2 M r + Q^2 + \frac{1}{l^2} r^4
\right) \partial_r X \right] \frac{1}{\sin \theta} \partial_\theta
\left( \sin \theta \partial_\theta X \right) - \frac{r^2}{w^2} X
\left( X^2 - 1 \right) = 0, \ee
or making the differentiation with respect to $r$ coordinate lead us
to
\ben \left( r^2 - 2 M r + Q^2 + \frac{1}{l^2} r^4 \right)
\partial_{rr} X + 2 \left( r - M + \frac{2}{l^2} r^3 \right)
\partial_r X \protect\\ \nonumber + \partial_{\theta \theta} X +
\frac{1}{\tan \theta} \partial_\theta X - \frac{r^2}{w^2} X \left( X^2
- 1 \right) = 0.
\label{sss}
\een
Next, we define the quantities
\be a \equiv r^2 - 2 M r + Q^2 + \frac{1}{l^2} r^4, ~~~~~~~ b \equiv r
- M + \frac{2}{l^2} r^3.
\label{abdef} 
\ee
It enables us to rewrite Eq.~(\ref{sss}) in the form as follows:
\be a \partial_{rr} X + 2 b \partial_r X + \partial_{\theta \theta} X
+ \frac{1}{\tan \theta} \partial_\theta X - \frac{r^2}{w^2} X \left(
X^2 - 1 \right) = 0.
\label{adsphi4a} 
\ee
Using the notation from our previous work~\cite{mr03}
Eq.~(\ref{adsphi4a}) can be rewritten in the finite differences
form.  It implies
\ben a \frac{X_{+0} - 2 X_{00} + X_{-0}}{(\Delta r)^2} + 2 b
\frac{X_{+0} - X_{-0}}{2 \Delta r} \protect\\ \nonumber + \frac{X_{0+}
- 2 X_{00} + X_{0-}}{(\Delta \theta)^2} + \frac{1}{\tan \theta}
\frac{X_{0+} - X_{0-}}{2 \Delta \theta} - \frac{r^2}{w^2} X_{00}
\left( X_{00}^2 - 1 \right) = 0,
\label{num1}
\een
or it can be rewritten in a more suitable form for our considerations.
Consequently, Eq.~(\ref{num1}) yields the result that:
\ben \left[ \frac{a}{(\Delta r)^2} + \frac{b}{\Delta r} \right] X_{+0}
+ \left[ \frac{a}{(\Delta r)^2} - \frac{b}{\Delta r} \right] X_{-0}
+ \left[ \frac{1}{(\Delta \theta)^2} + \frac{1}{2 \Delta \theta \tan
\theta} \right] X_{0+} \protect\\ \nonumber + \left[ \frac{1}{(\Delta
\theta)^2} - \frac{1}{2 \Delta \theta \tan \theta} \right] X_{0-} +
\left[ -\frac{2 a}{(\Delta r)^2} - \frac{2}{(\Delta \theta)^2} -
\frac{r^2}{w^2} (X_{00}^2 - 1) \right] X_{00} = 0.
\label{adsphi4b}
\een
We solve the equation (\ref{adsphi4b}) using simultaneously
over-relaxation (SOR) method~\cite{pre92}.  A general form of the
second-order elliptic equation, finite differenced, yields
\be A x_{+0} + B x_{-0} + C x_{0+} + D x_{0-} + E x_{00} = F.
\label{num2}
\ee
Equation~(\ref{adsphi4b}) can be brought to the form of relation
(\ref{num2}) with the following coefficients for the SOR method for
our problem:
\ben A = \frac{a}{(\Delta r)^2} + \frac{b}{\Delta r}, \\ B =
\frac{a}{(\Delta r)^2} - \frac{b}{\Delta r}, \\ C = \frac{1}{(\Delta
\theta)^2} + \frac{1}{2 \Delta \theta \tan \theta_k}, \\ D =
\frac{1}{(\Delta \theta)^2} - \frac{1}{2 \Delta \theta \tan \theta_k},
\\ E = -\frac{2}{(\Delta r)^2} - \frac{2}{(\Delta \theta)^2} -
\frac{r^2}{w^2} (X_{00}^2 - 1) \\ F = 0.  \een

\subsubsection{Boundary condition on the event horizon of black hole}
On the event horizon of the black hole under consideration $a=0$ and
it can be verified that equation of motion (\ref{motion}) takes the
form
\be 2 \left( r_{EH} - M + \frac{2}{l^2} r_{EH}^3 \right) \partial_r X
\vert_{r=r_{EH}} + \partial_{\theta \theta} X + \frac{1}{\tan \theta}
\partial_\theta X - \frac{r_{EH}^2}{w^2} X \left( X^2 - 1 \right) = 0,
\ee
which yields the finite difference equation for our problem
\be 2 b_{EH} \frac{X_{+0} - X_{00}}{\Delta r} + \frac{X_{0+} - 2
X_{00} + X_{0-}}{(\Delta \theta)^2} + \frac{1}{\tan \theta}
\frac{X_{0+} - X_{0-}}{2 \Delta \theta} - \frac{r_{EH}^2}{w^2} X_{00}
\left( X_{00}^2 - 1 \right) = 0.
\label{num3}
\ee
It can seen that by virtue of relation (\ref{num3}) one obtains the
following:
\ben \frac{2 b_{EH}}{\Delta r} X_{+0} + \left[ \frac{1}{2 \Delta
\theta \tan \theta} \right] X_{0+} + \left[ \frac{1}{(\Delta
\theta)^2} - \frac{1}{2 \Delta \theta \tan \theta} \right] X_{0-} +
\protect\\ \nonumber \left[ -\frac{2 a}{(\Delta r)^2} -
\frac{2}{(\Delta \theta)^2} - \frac{r_{EH}^2}{w^2} (X_{00}^2 - 1)
\right] X_{00} = 0, \een
where we have denoted by $b_{EH}$ the following expression:
\be b_{EH} \equiv r_{EH} - M + \frac{2}{l^2} r_{EH}^3.
\label{beh}
\ee
Now, the coefficients for the SOR method are as follows:
\ben A_0 = \frac{2 b_{EH}}{\Delta r}, \\ B_0 = 0, \\ C_0 =
\frac{1}{(\Delta \theta)^2} + \frac{1}{2 \Delta \theta \tan \theta} \\
D_0 = \frac{1}{(\Delta \theta)^2} - \frac{1}{2 \Delta \theta \tan
\theta}, \\ E_0 = -\frac{2 b_{EH}}{\Delta r} - \frac{2}{(\Delta
\theta)^2} - \frac{r_{EH}^2}{w^2} (X_{00}^2 - 1), \\ F_0 = 0.  \een

\subsubsection{Extremal black hole}
For the extremal black hole both $a=0$ and $b=0$ on the event horizon
of the black hole.  The field equation on the horizon decouples from
the rest of the space and takes the form
\be \partial_{\theta \theta} X + \frac{1}{\tan \theta} \partial_\theta
X - \alpha X \left( X^2 - 1 \right) = 0,
\label{adsphi4c} 
\ee
where $\alpha \equiv r_{EH}^2/w^2$.

As was shown in Ref.~\cite{mr04} for $\alpha<2$ we have a field
expulsion from the extremal RN black hole and one can set $X=0$ as a
boundary condition on the event horizon.  For $\alpha>2$ equation
(\ref{adsphi4c}) must be solved prior to the calculation for the rest
of the grid.  For the case of the extremal black hole we shall adopt
two point boundary relaxation method~\cite{pre92}.

\subsection{Cosmological AdS black hole -- Sine-Gordon potential}
For the sine-Gordon potential the only difference in the SOR method
are the $E$ and $F$ coefficients, which in this case take the form
\ben E = -\frac{2 a}{(\Delta r)^2} - \frac{2}{(\Delta \theta)^2},\\ F
= -\frac{r^2}{w^2} \sin X_{00}.  \een
On the other hand, on the event horizon we have the following
coefficients:
\ben E_0 = -\frac{2 b_{EH}}{(\Delta r)^2} - \frac{2}{(\Delta
\theta)^2}, \\ F_0 = -\frac{r_{EH}^2}{w^2} \sin X_{00}.  \een

\subsection{Anti-Nariai black holes}
\begin{figure*}
\begin{center}
\leavevmode
\hbox{%
\includegraphics[width=240pt]{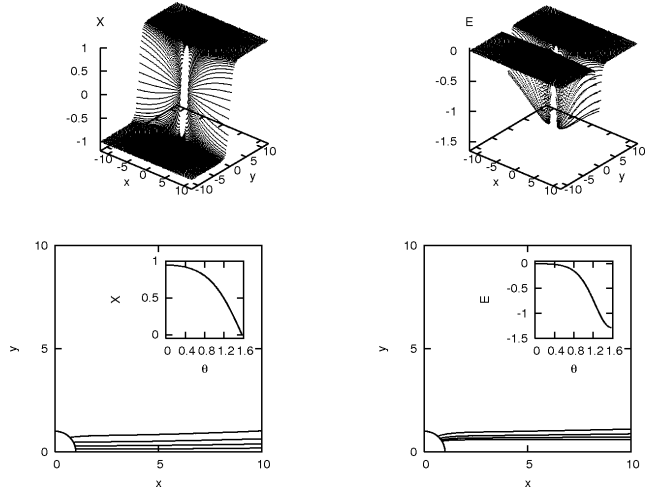}}
\end{center}
\caption{The field X (left panels) and the quantity E (right panels)
  for the $\phi^4$ potential and the cosmological AdS black
  hole. Isolines on bottom panels are drawn for $0.2$, $0.4$, $0.6$
  and $0.8$ for the field X and for $-0.1$, $-0.2$, $-0.3$ and $-0.4$
  for the energy. Inlets in bottom plots show the value of the fields
  on the black hole horizon. Black hole has $M=1.0$, $Q=0.1$, $l=1.0$
  and the domain width is $w=1.0$.}
\label{fig:plot1}
\end{figure*}

\begin{figure*}
\begin{center}
\leavevmode
\hbox{%
\includegraphics[width=240pt]{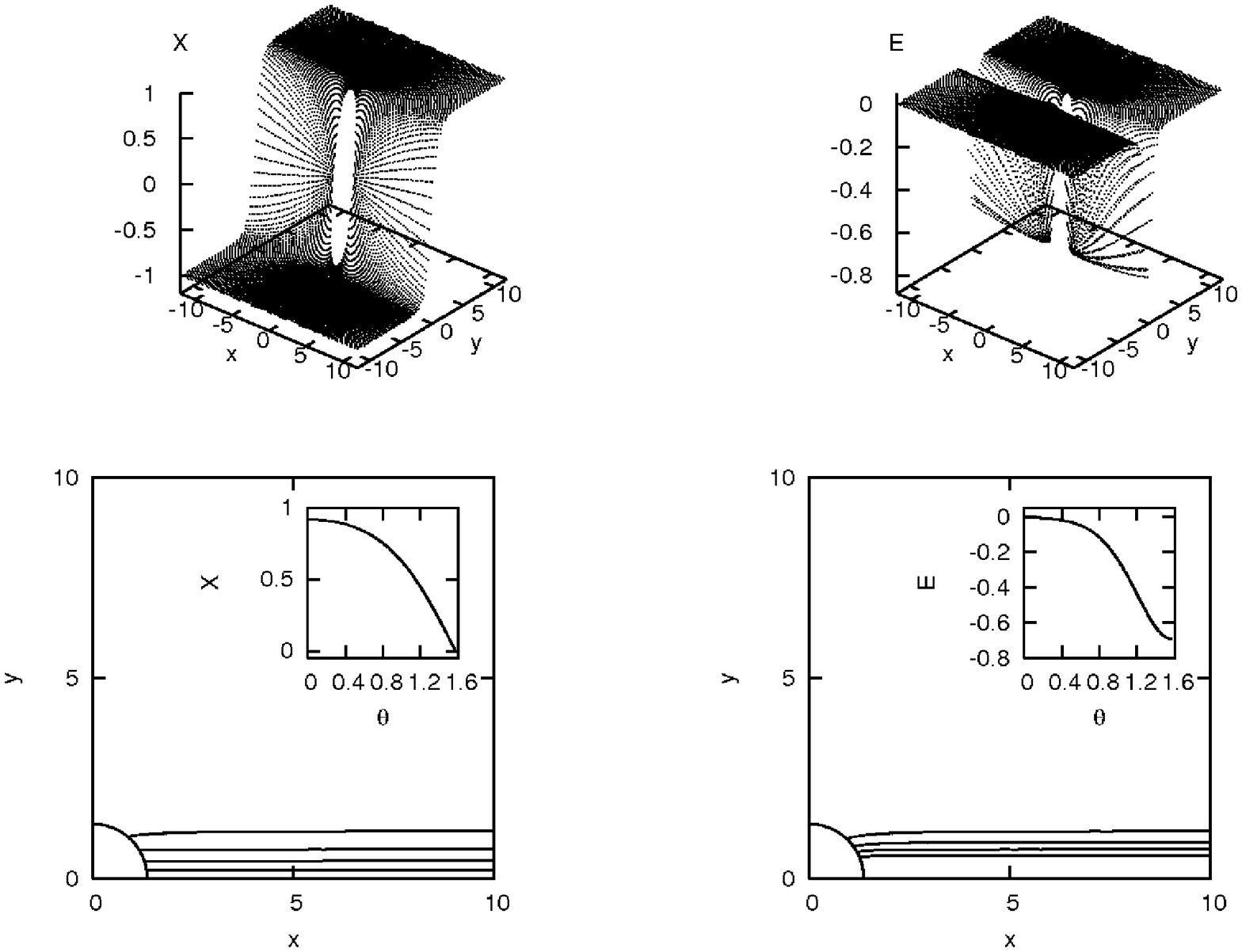}}
\end{center}
\caption{Same as Fig.~\ref{fig:plot1}, but for $l=2.0$}
\label{fig:plot2}
\end{figure*}

\begin{figure*}
\begin{center}
\leavevmode
\hbox{%
\includegraphics[width=240pt]{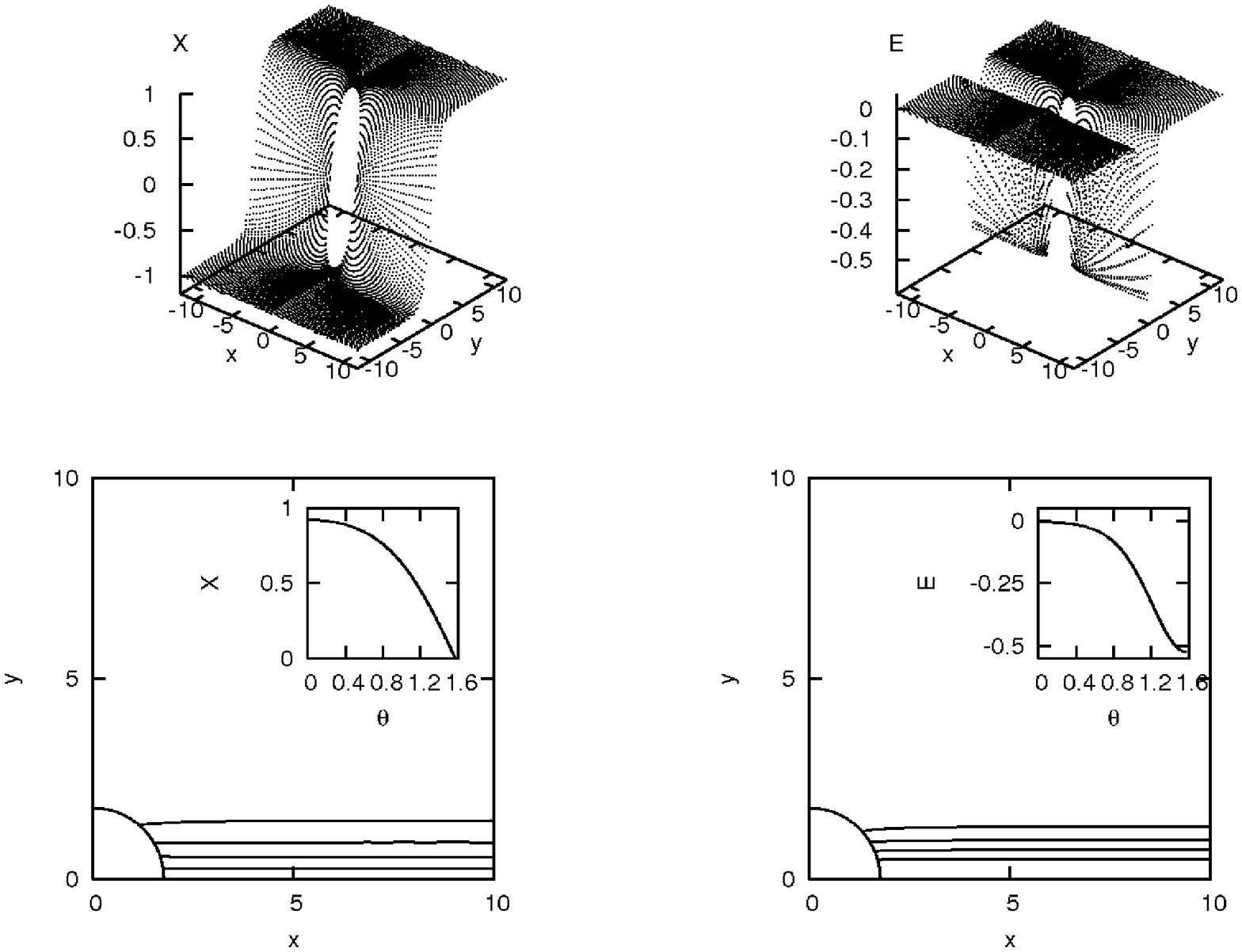}}
\end{center}
\caption{Same as Fig.~\ref{fig:plot1}, but for $l=5.0$}
\label{fig:plot5}
\end{figure*}

\begin{figure*}
\begin{center}
\leavevmode
\hbox{%
\includegraphics[width=240pt]{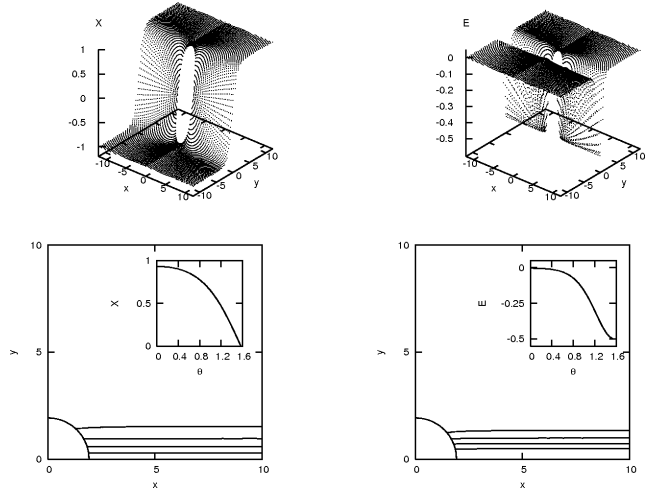}}
\end{center}
\caption{Same as Fig.~\ref{fig:plot1}, but for $l=10.0$}
\label{fig:plot10}
\end{figure*}

Now we pay attention to the anti-Nariai black holes.  The metric can
be obtained by Ginsparg-Perry procedure~\cite{gp83} from near-extreme
AdS black hole.  The AdS black hole that generates anti-Nariai one is
of hyperbolic type.  The anti-Nariai solution can be get when the the
black hole horizon approaches the cosmological horizon.  Its line
element may be written in the form as~\cite{dl03}
\be ds^2 = - N(R) dT^2 + {dR^2 \over N(R)} + {\cR}_{0}^2 (d \theta^2 +
\sinh \theta ^2 d \phi^2), \ee
where $N(R) = - 1 + {K_{0} \over {\cR}_{0}^2 }R^2$, and
\be \Lambda = - {1 + K_{0}\over 2 {\cR}_{0}^2}, \qquad q^2 = {K_{0} -
1 \over 2} {\cR}_{0}^2.  \ee
For anti-Nariai case one has to replace the definitions (\ref{abdef})
and (\ref{beh}) with
\be a \equiv R_0^2 N(R), \qquad b \equiv R_0^2 N'(R),
\label{abdefan}
\ee
and
\be b_{EH} \equiv R_0^2 N'(R_{EH}).  \ee
Then the SOR coefficient are the same as in the previous case with
only one exception of
\be E = -\frac{2}{(\Delta r)^2} - \frac{2}{(\Delta \theta)^2} -
\frac{R_0^2}{w^2} (X_{00}^2 - 1), \qquad E_0 = -\frac{2 b_{EH}}{\Delta
r} - \frac{2}{(\Delta \theta)^2} - \frac{R_0^2}{w^2} (X_{00}^2 - 1),
\ee
for $\phi^4$ potential, and
\be F = F_0 = -\frac{R_0^2}{w^2} \sin X_{00}, \ee
for sine-Gordon potential.
To conclude this section, we comment briefly on results of our
numerical investigations.  First we have studied the behaviour of the
black hole domain wall system due to changes of cosmological
parameter.  Next we pay special attention to extreme black holes and
domain walls. These systems are of special interests because of the
possibility of expulsion of the scalar field from the black hole (the
so-called {\it Meissner effetct}).  Namely, Figure~\ref{fig:plot1}
represents the results of numerical integrations of equations of
motion for the AdS black hole with $\phi^4$-potential and the mass $M
= 1.0$, whereas the cosmological parameter $l = 1$. The width of the
domain wall was taken to be $w = 1$. We depicted the values of the
scalar field $X$ and the quantity $E = {T_{t}{}{}^{t} \over \la
\eta^4}$. In Figs.~\ref{fig:plot2}-\ref{fig:plot10} we take into
account the growth of the cosmological parameter.  Namely, in
Fig.~\ref{fig:plot2} we have $l = 2$, while in Fig.~\ref{fig:plot5}
one has $l = 5$ and in Fig.~\ref{fig:plot10} $l = 10$.  The remaining
parameters are the same as in Fig.~\ref{fig:plot1}.

\begin{figure*}
\begin{center}
\leavevmode
\hbox{%
\includegraphics[width=240pt]{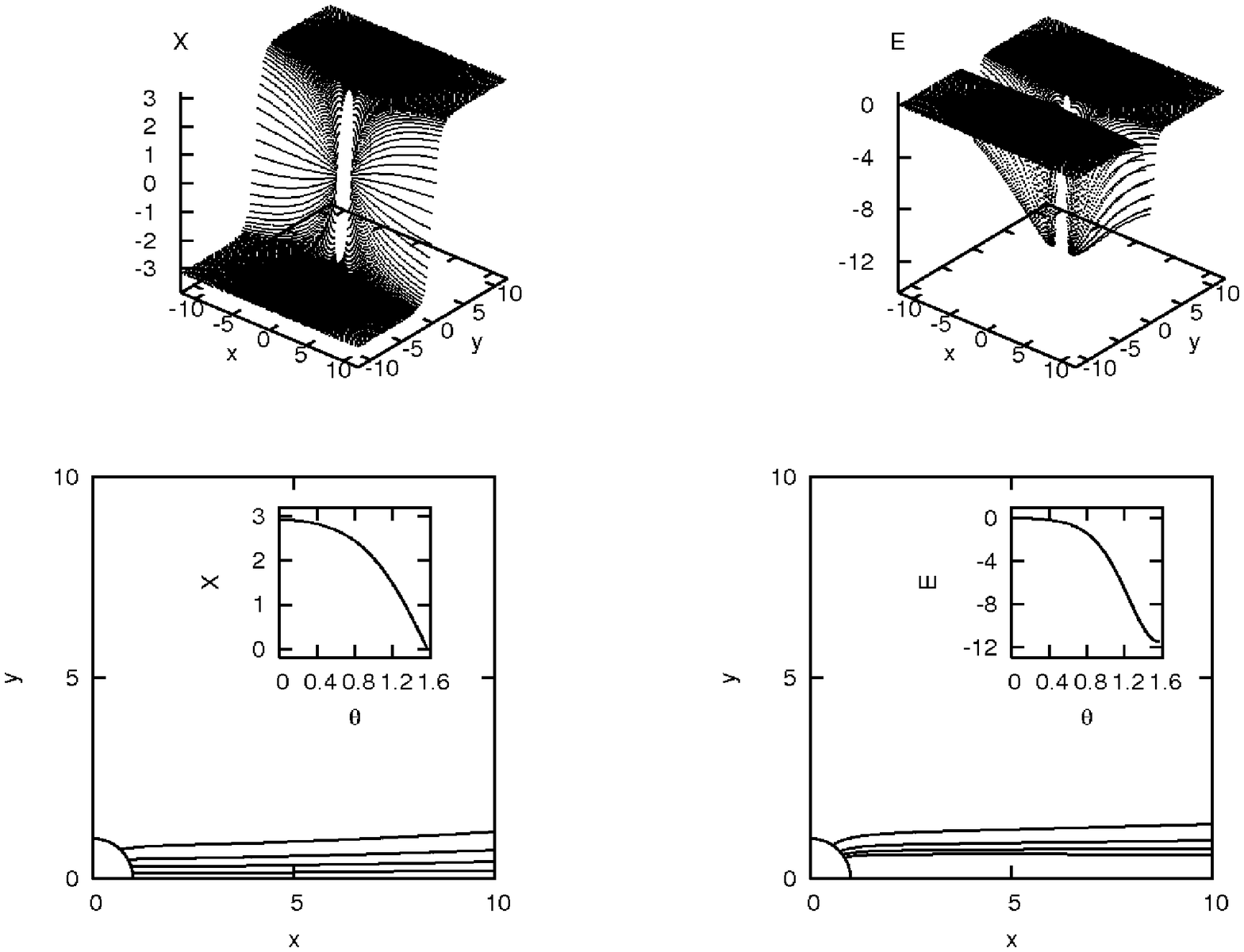}}
\end{center}
\caption{The field X (left panels) and the quantity E (right panels)
  for the sine-Gordon potential and the cosmological AdS black
  hole. Isolines on bottom panels are drawn for $0.2\pi$, $0.4\pi$, $0.6\pi$
  and $0.8\pi$ for the field X and for $-0.5$, $-1.5$, $-2.5$ and $-3.5$
  for the energy. Inlets in bottom plots show the value of the fields
  on the black hole horizon. Black hole has $M=1.0$, $Q=0.0$, $l=1.0$
  and the domain width is $w=1.0$.}
\label{fig:plot1s}
\end{figure*}

\begin{figure*}
\begin{center}
\leavevmode
\hbox{%
\includegraphics[width=240pt]{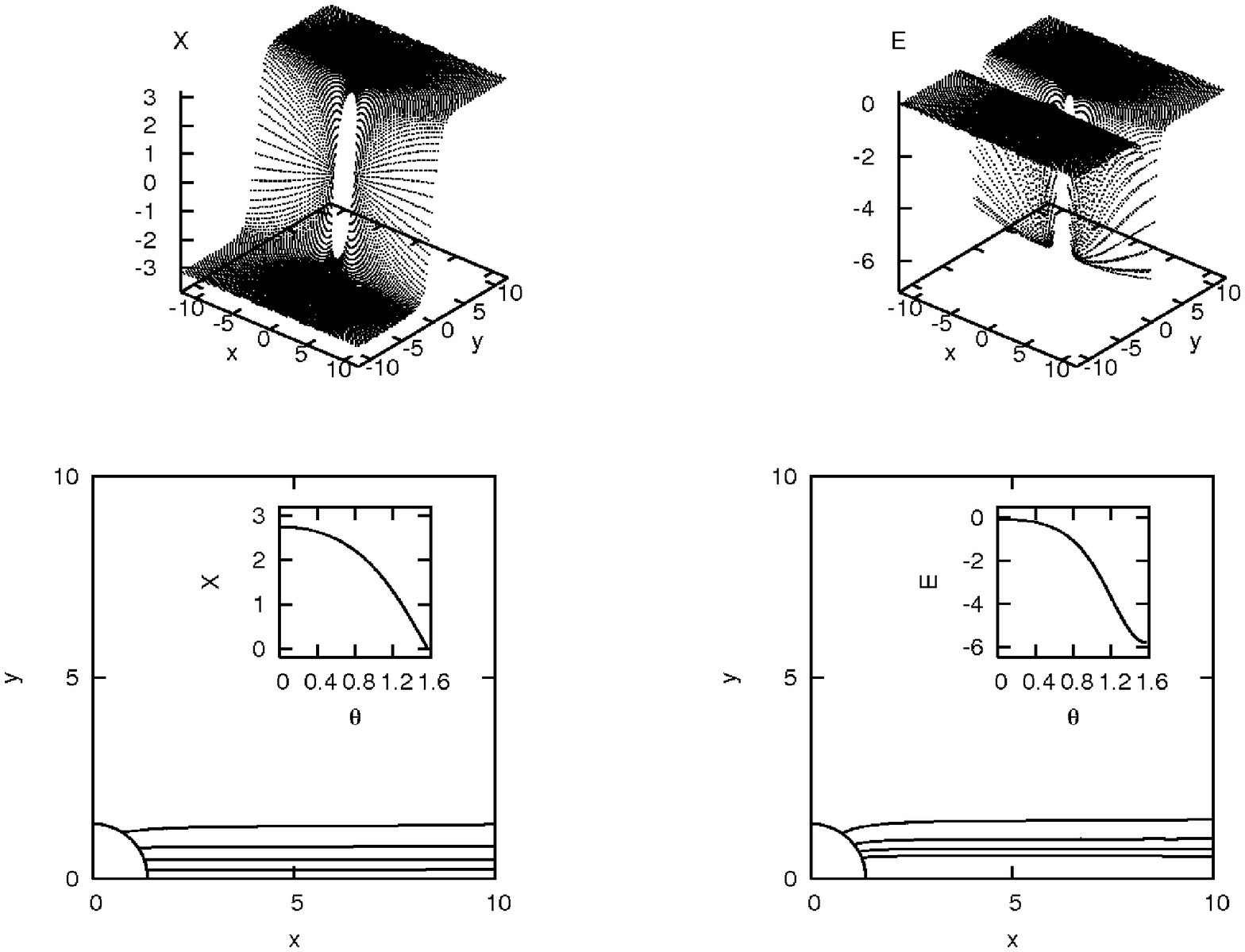}}
\end{center}
\caption{Same as Fig.~\ref{fig:plot1s}, but for $l=2.0$}
\label{fig:plot2s}
\end{figure*}

\begin{figure*}
\begin{center}
\leavevmode
\hbox{%
\includegraphics[width=240pt]{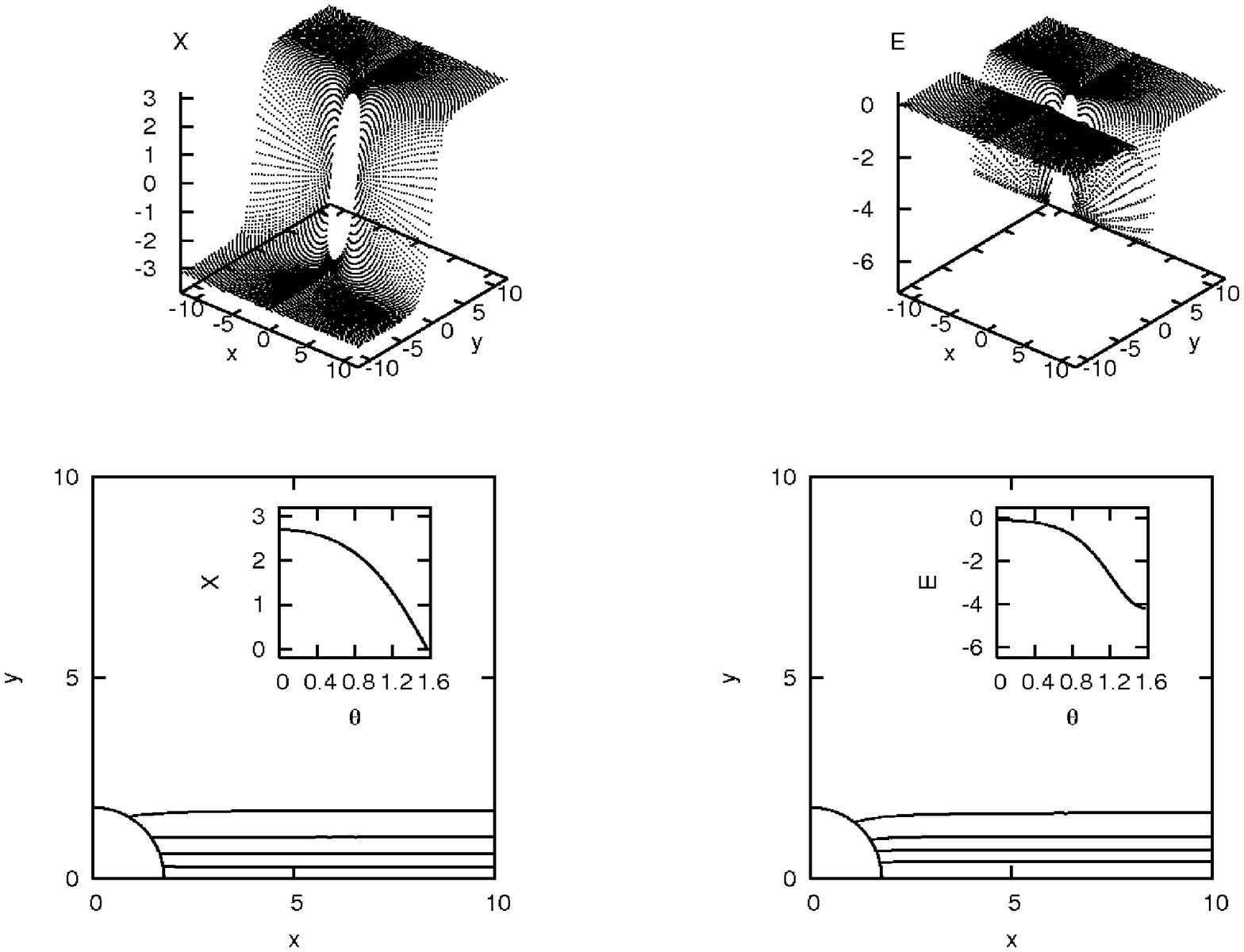}}
\end{center}
\caption{Same as Fig.~\ref{fig:plot1s}, but for $l=5.0$}
\label{fig:plot5s}
\end{figure*}

\begin{figure*}
\begin{center}
\leavevmode
\hbox{%
\includegraphics[width=240pt]{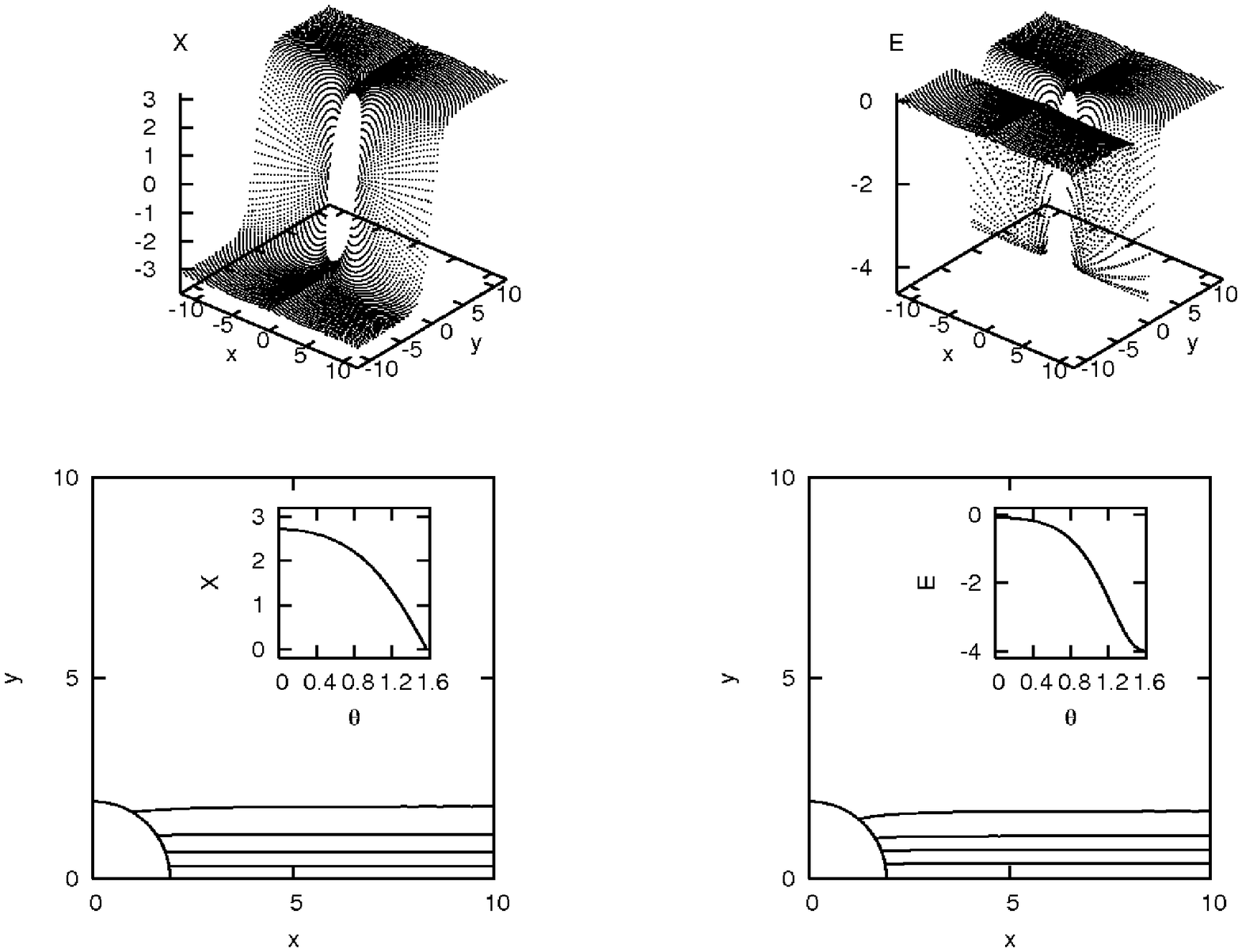}}
\end{center}
\caption{Same as Fig.~\ref{fig:plot1s}, but for $l=10.0$}
\label{fig:plot10s}
\end{figure*}

\begin{figure}
\begin{center}
\leavevmode
\hbox{%
\includegraphics[width=240pt]{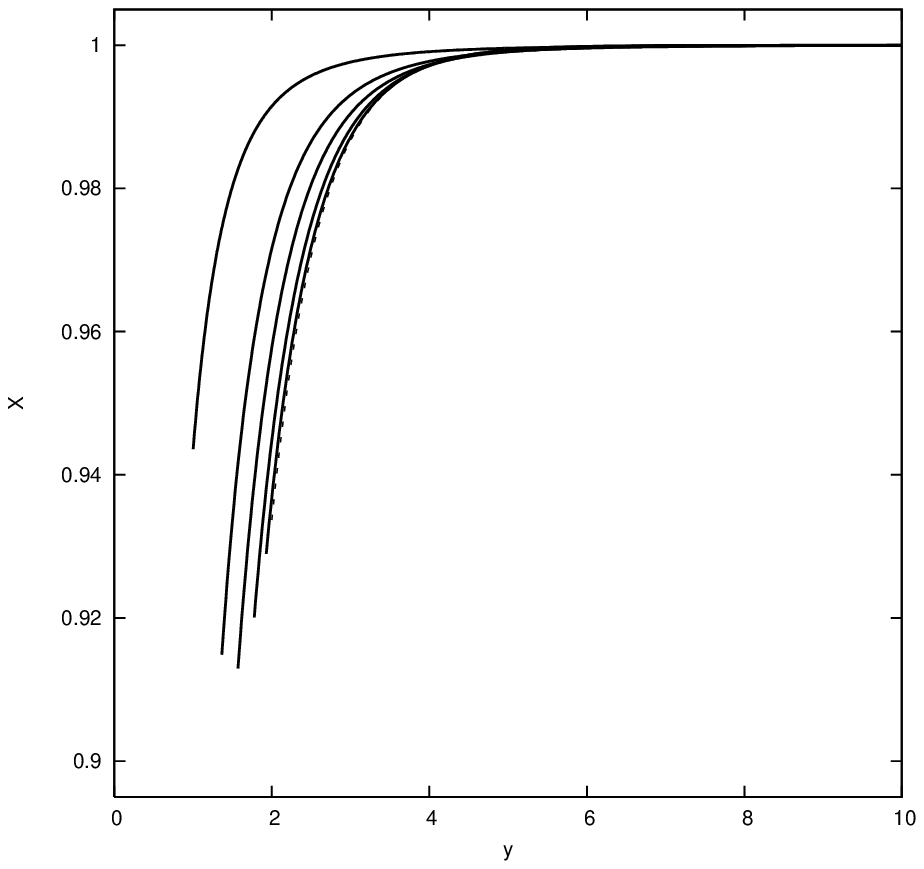}}
\end{center}
\caption{Values of the field X on the black hole axis for
  $l=1,2,3,5,10$ -- solid curves from left to right,
  respectively. Dashed line represents the solution for $l=\inf$. Rest
  of the parameters are the same as in Figure~\ref{fig:plot1}.}
\label{fig:zax}
\end{figure}

\begin{figure}
\begin{center}
\leavevmode
\hbox{%
\includegraphics[width=240pt]{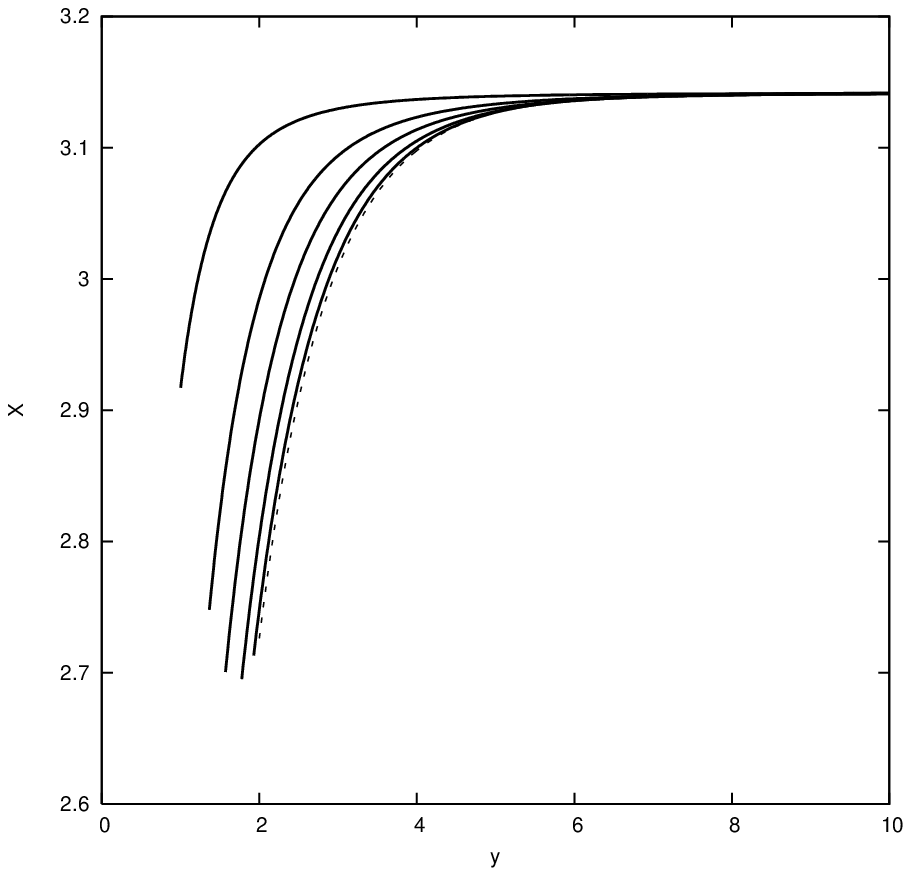}}
\end{center}
\caption{Same as Fig.~\ref{fig:zax}, but for sine-Gordon potential.}
\label{fig:zaxs}
\end{figure}

Figures~\ref{fig:plot1s}-\ref{fig:plot10s} depict the solutions of
equations of motion for scalar field $X$ and show changes of $E$, for
AdS black hole with sine-Gordon potential.  The mass of the black hole
under consideration one has equal to $M = 1.0$ and the cosmological
parameters are chosen respectively to be $1$, $2$, $5$ and $10$.  The
width of the domain wall was taken to be $1$.
On the other hand, Figures~\ref{fig:zax}-\ref{fig:zaxs} show the behaviour
of the scalar field $X$ on the black hole axis both for $\phi^4$ and
sine-Gordon potentials.

\begin{figure}
\begin{center}
\leavevmode
\hbox{%
\includegraphics[width=240pt]{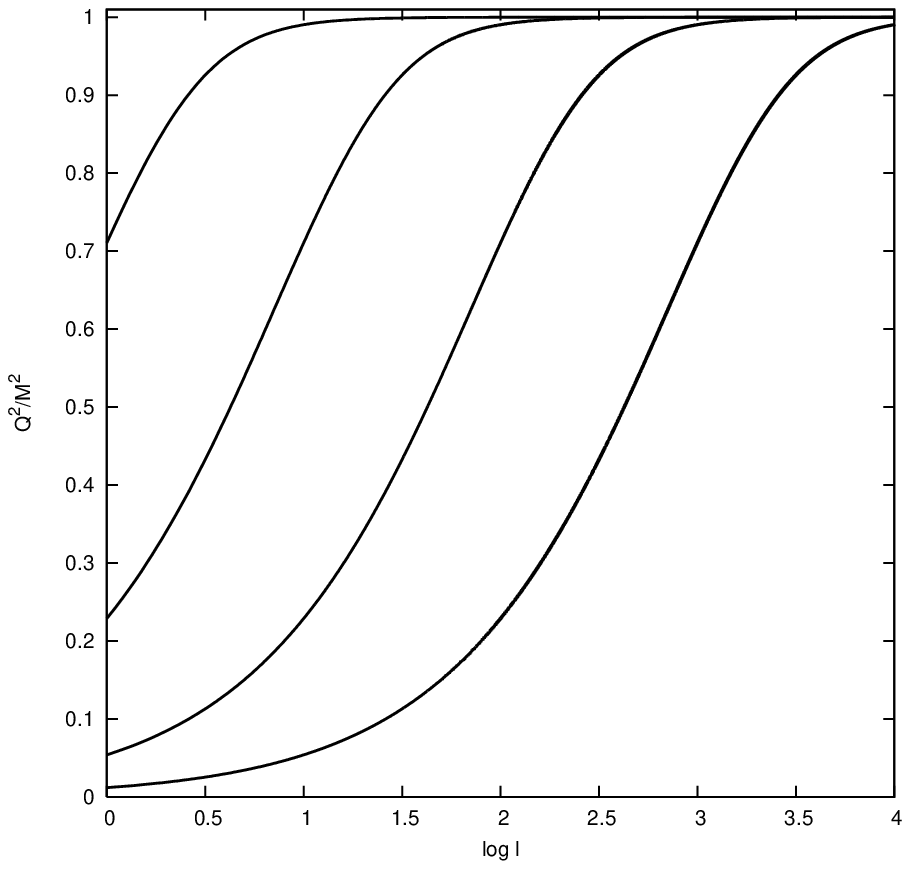}}
\end{center}
\caption{$Q^2/M^2$ as a function of the cosmological parameter $l$ for
  extreme black hole.  The curves are for $M=1,10,100,1000$, from left
  to right.}
\label{fig:qmax}
\end{figure}

Figure~\ref{fig:qmax} presents values of the ratio $Q^2/M^2$ for which
the black hole under considerations becomes extreme.  We consider it
as a function of the cosmological parameter $l$ and plots are made for
different values of the black hole masses, i.e., $M =
1,~10,~100,~1000$.

\begin{figure*}
\begin{center}
\leavevmode
\hbox{%
\includegraphics[width=240pt]{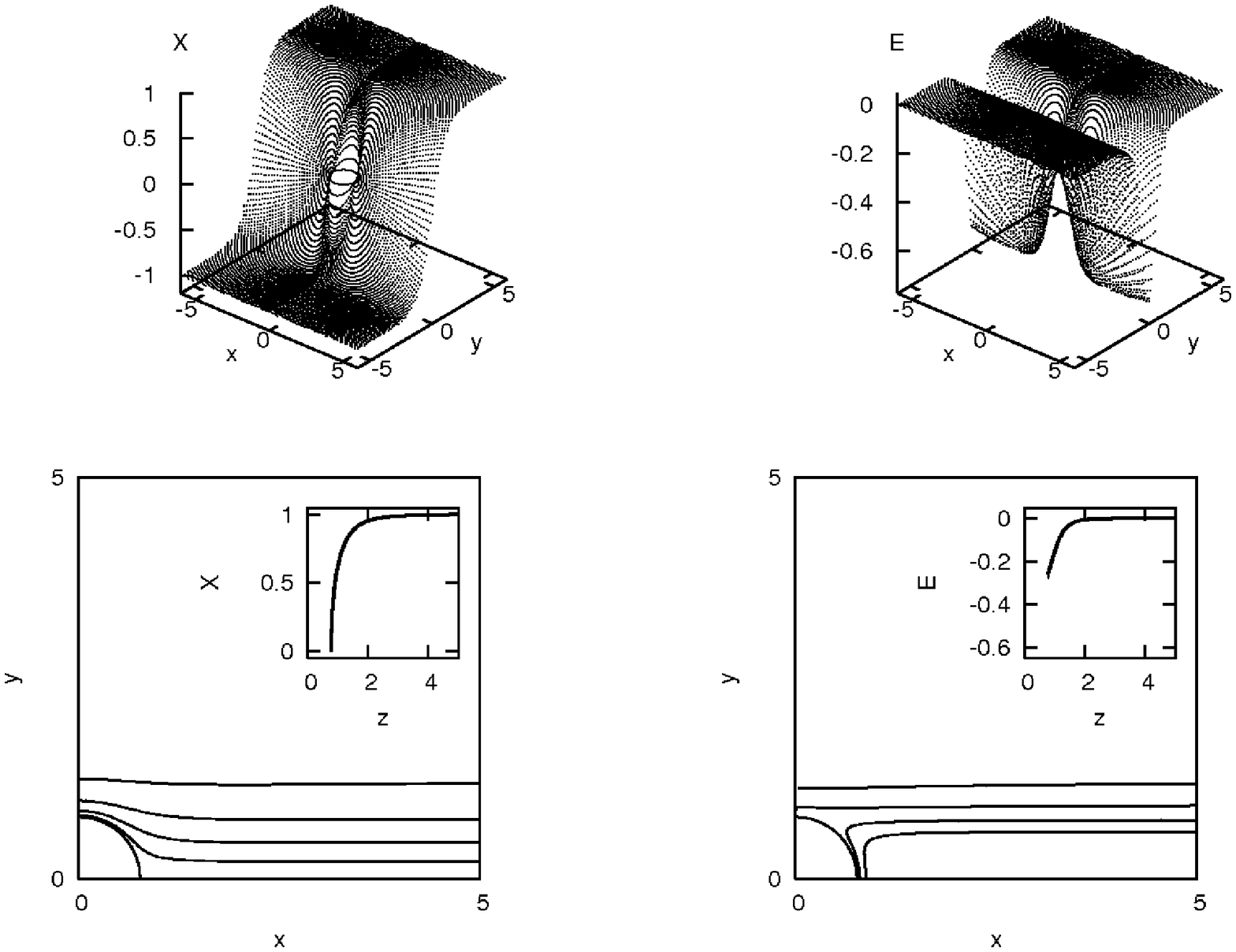}}
\end{center}
\caption{The field X (left panels) and E (right panels) for the
  $\phi^4$ potential and the extremal cosmological AdS black hole.
  Isolines on bottom panels are drawn for $0.2$, $0.4$, $0.6$ and
  $0.8$ for the field X and for $-0.5$, $-1.5$, $-2.5$ and $-3.5$ for
  E.  Inlets in bottom plots show the value of the fields on the black
  hole horizon.  Black hole has $M=1.0$, the cosmological parameter
  $l=1.0$ and $\alpha \approx 0.6$.}
\label{fig:plot2a}
\end{figure*}

\begin{figure*}
\begin{center}
\leavevmode
\hbox{%
\includegraphics[width=240pt]{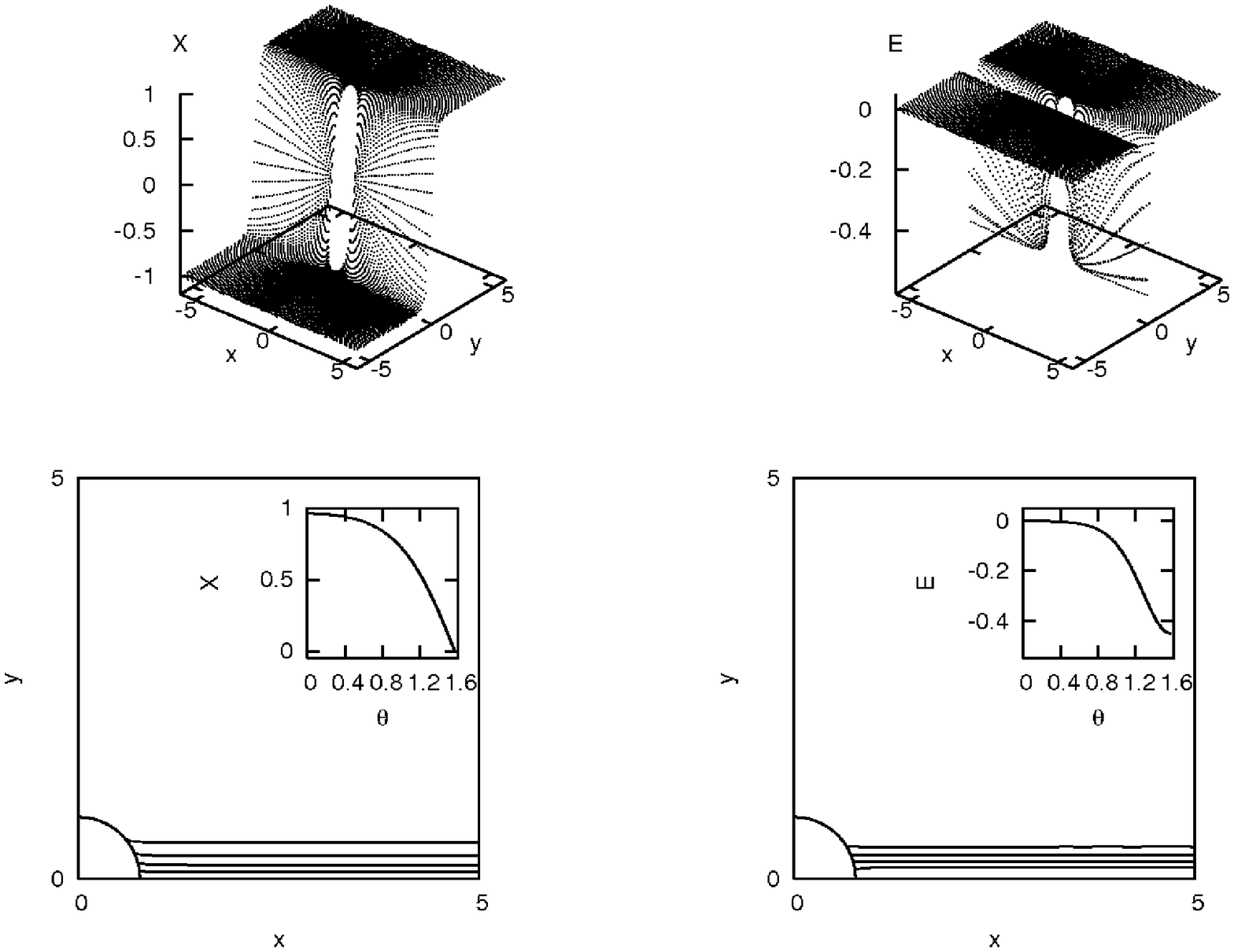}}
\end{center}
\caption{Same as Fig.~\ref{fig:plot2a}, but for $\alpha \approx 6.6$.}
\label{fig:plot2b}
\end{figure*}

\begin{figure*}
\begin{center}
\leavevmode
\hbox{%
\includegraphics[width=240pt]{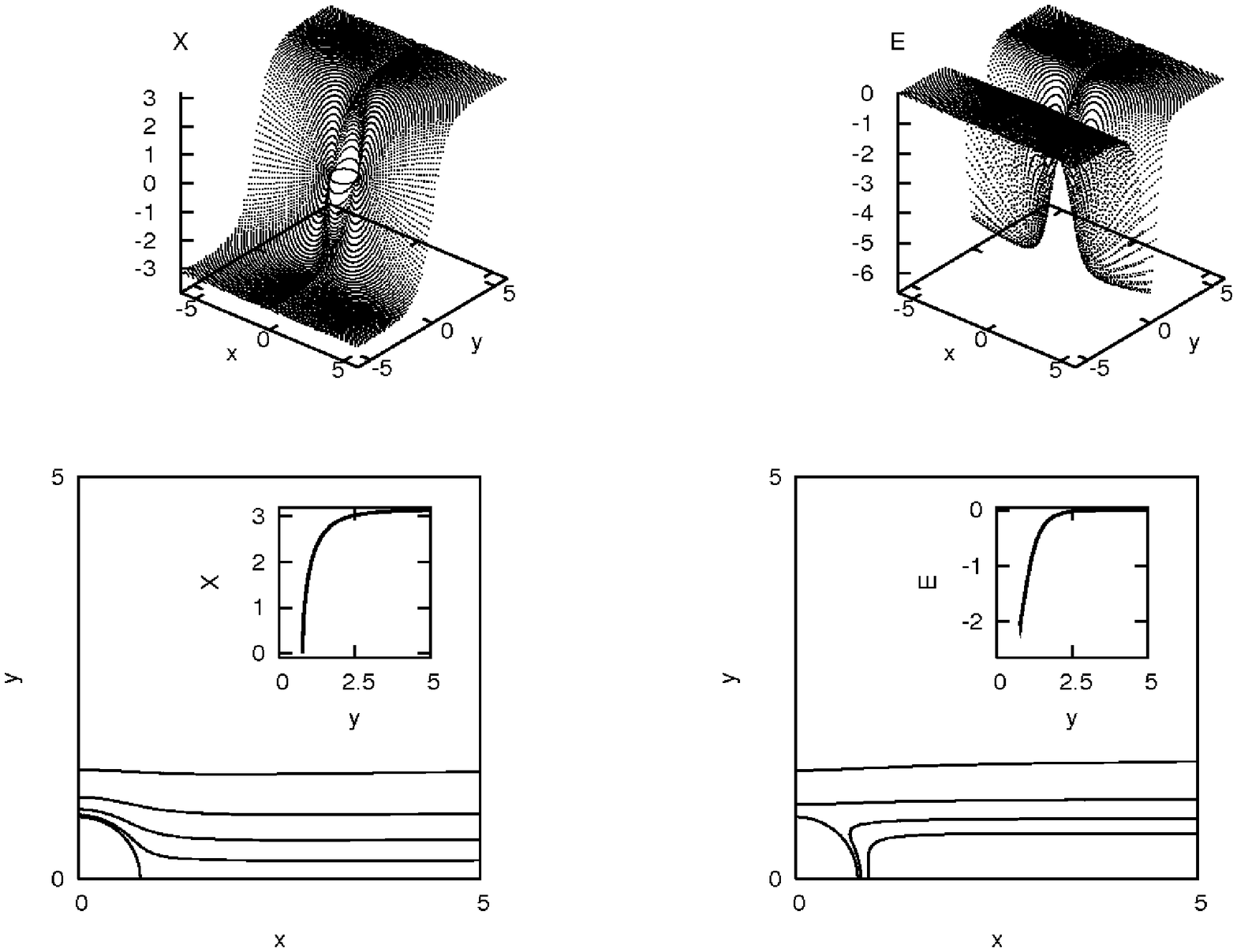}}
\end{center}
\caption{Same as Fig.~\ref{fig:plot2a}, but for sine-Gordon potential.}
\label{fig:plot2as}
\end{figure*}

\begin{figure*}
\begin{center}
\leavevmode
\hbox{%
\includegraphics[width=240pt]{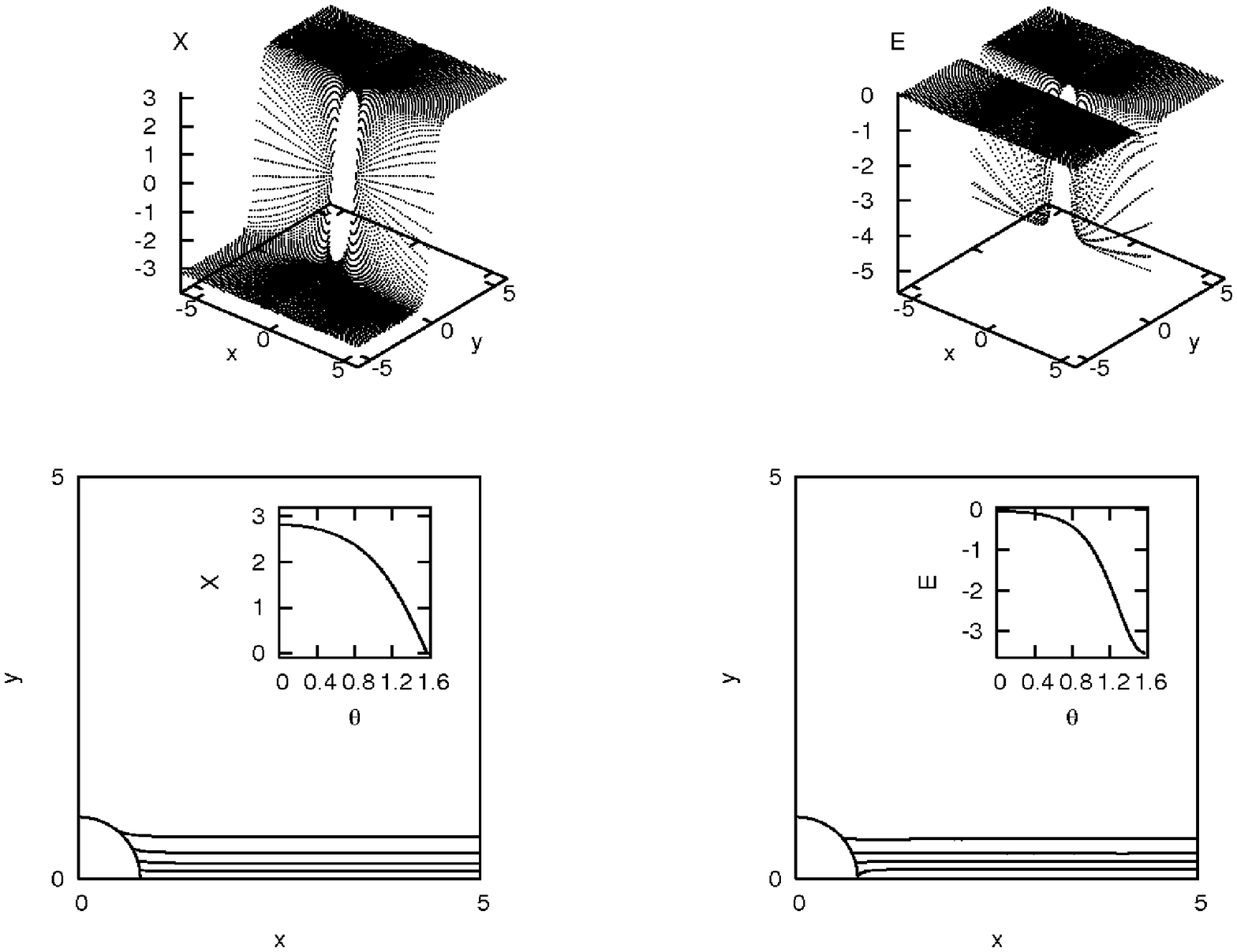}}
\end{center}
\caption{Same as Fig.~\ref{fig:plot2as}, but for $\alpha \approx 6.6$.}
\label{fig:plot2bs}
\end{figure*}

In Figs.~\ref{fig:plot2a}-\ref{fig:plot2bs} we plotted the values of
$X$ and the parameter $E$ for extremal black holes.  It revealed that
the field expulsion takes place only when the value of $\alpha$
parameter is smaller than $2.0$.  The same situation we ancountered in
the case of RN black holes~\cite{mr04}.  On the contrary, in the case
of extremal dilaton black hole domain wall system one gets always
expulsion of the scalar field~\cite{rog01,mr03}.  It was revealed that
the simplest generalization of Einstein-Maxwell theory by adding a
massless dilaton field dramatically changes the structure and
properties of extremal black holes.  Figure~\ref{fig:plot2a} depicts
the value of the scalar field $X$ and $E$ for $\alpha \sim 0.6$ and
for $\phi^4$-potential.  In Fig.~\ref{fig:plot2b} we change the value
of $\alpha$.  Now it is equal to $6.6$.  The results depicted in
Figs.~\ref{fig:plot2as}-\ref{fig:plot2bs} are valid for sine-Gordon
potential and $\alpha = 0.6$ and $6.6$, respectively.

\begin{figure*}
\begin{center}
\leavevmode
\hbox{%
\includegraphics[width=120pt]{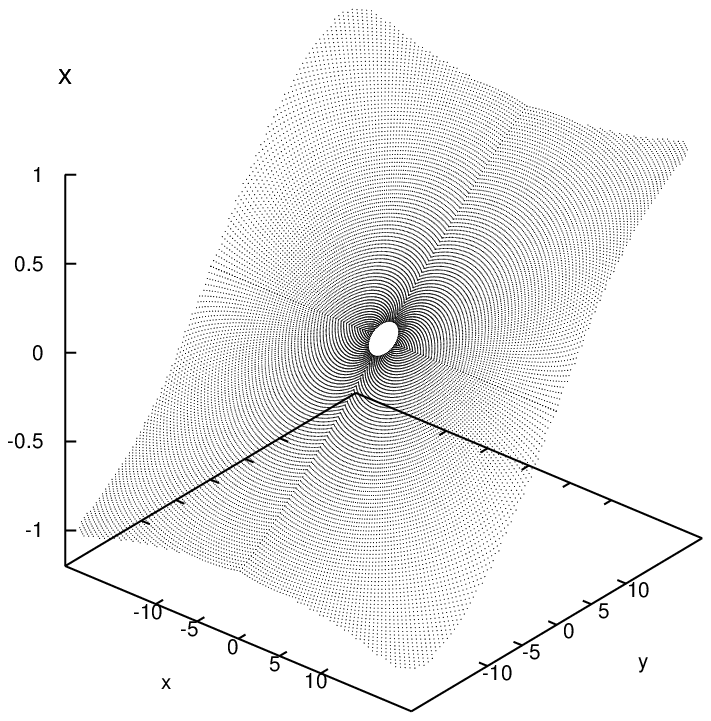}}
\hskip 1cm
\hbox{%
\includegraphics[width=120pt]{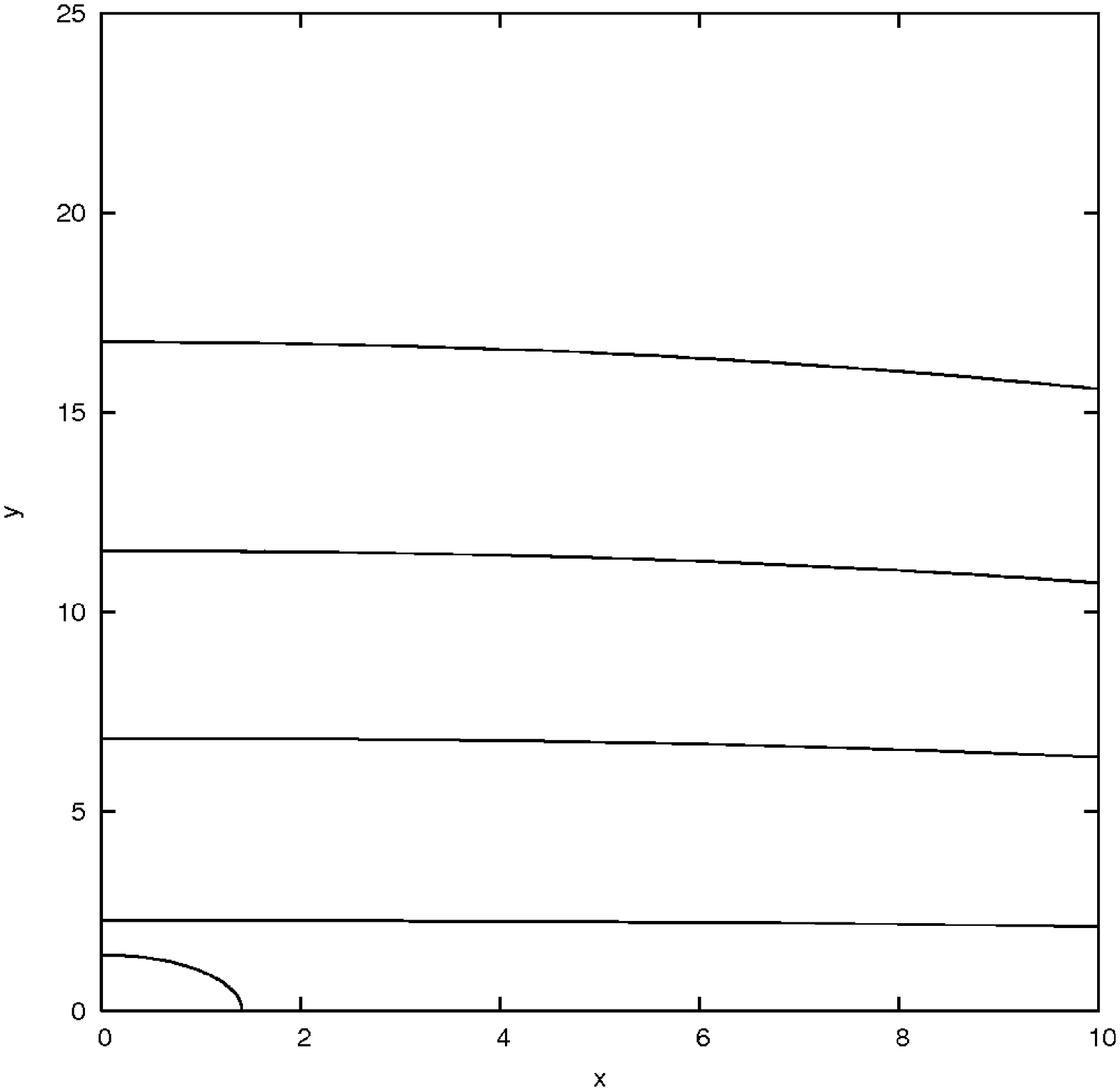}}
\end{center}
\caption{X field configuration for anti-Nariai black hole $R_{0} =1.0$
  and $K_{0}= 1/2$ for $\phi^4$ potential. Left panel shows the value
  of the field, and right panel presents $0.2, 0.4, 0.6, 0.8$
  contours.}
\label{fig:plotan}
\end{figure*}

\begin{figure*}
\begin{center}
\leavevmode
\hbox{%
\includegraphics[width=120pt]{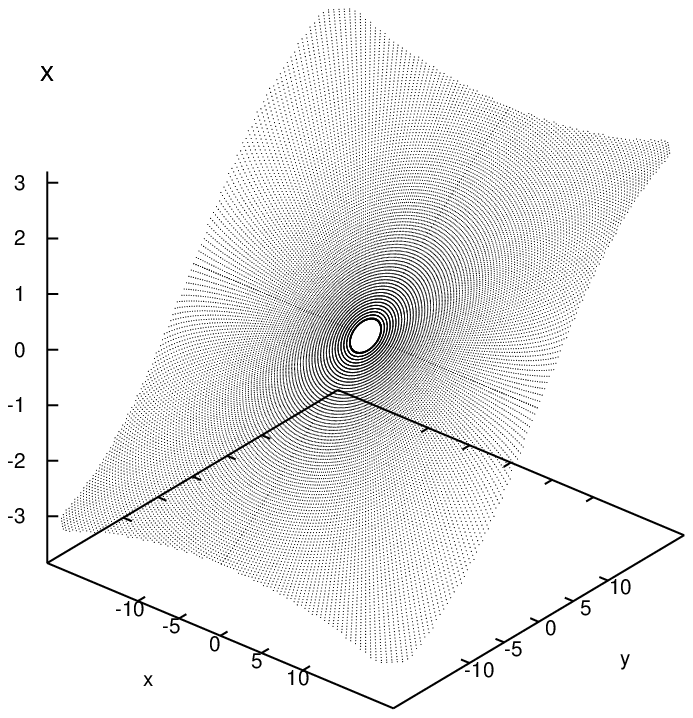}}
\hskip 1cm
\hbox{%
\includegraphics[width=120pt]{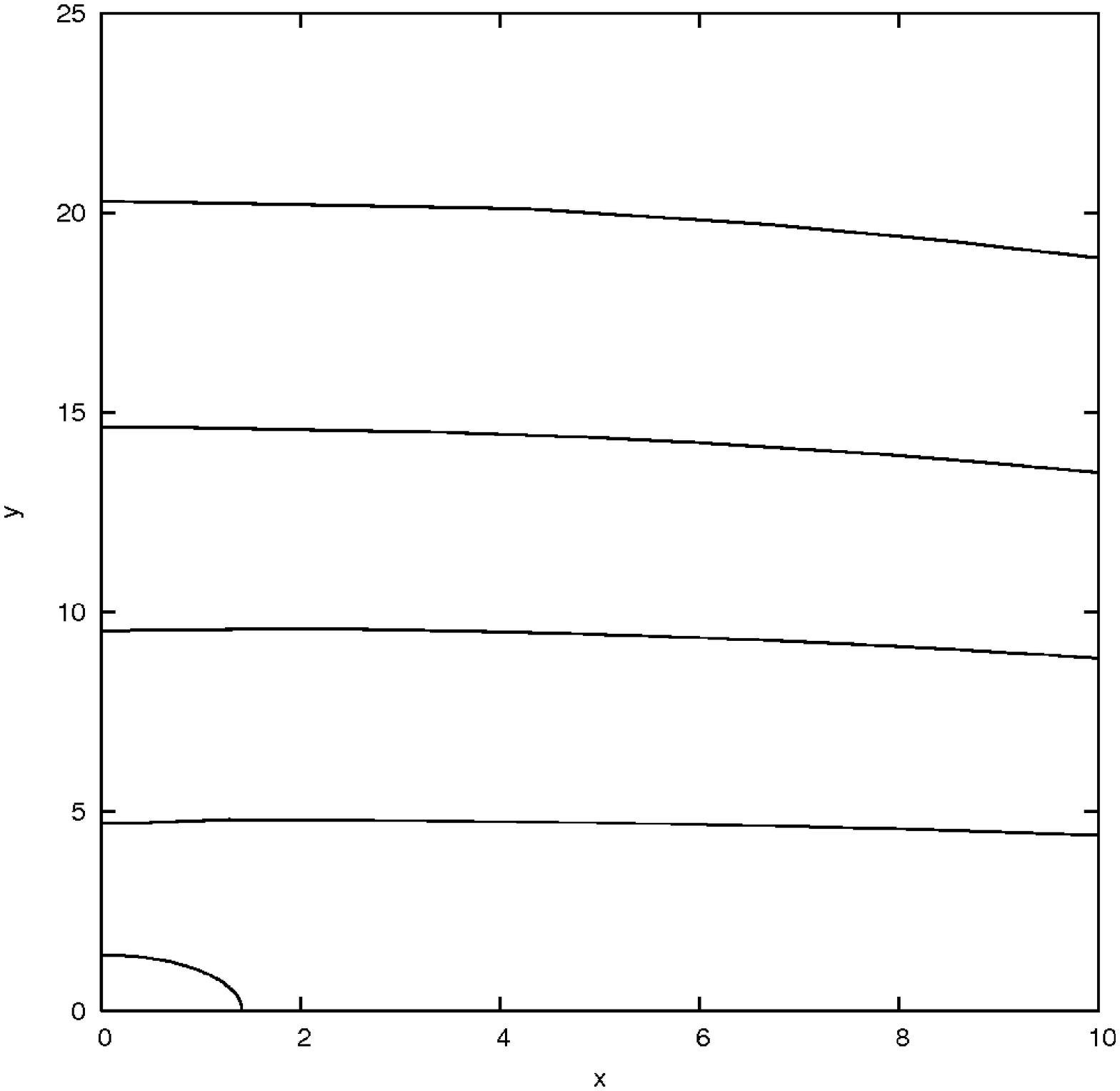}}
\end{center}
\caption{Same as Fig.~\ref{fig:plotan}, but for sine-Gordon
  potential. Contours on the right panel are for $0.2\pi, 0.4\pi,
  0.6\pi, 0.8\pi$.}
\label{fig:plotans}
\end{figure*}

In Figs.~\ref{fig:plotan}-\ref{fig:plotans} we show the $X$ and $E$
values for anti-Nariai black hole with $R_{0} = 1$ and $K_{0} = 1/2$
both for $\phi^4$ and sine-Gordon potentials.  In Ref.~\cite{rog04}
the charged Nariai and anti-Nariai domain wall systems were studied.
Due to the complication of equations of motion the simple arguments
concerning the problem of expulsion of scalar field were given.
However, the present analysis reveals the fact that there is no
expulsion for the anti-Nariai case.  Nevertheless, in the vicinity of
the event horizon one obtains very small value of the field $X$.

\section{\label{sec:conc}Conclusions}
Our considerations were devoted to studies of AdS black hole domain
wall system.  As in our previous works \cite{mr03,mr04} the domain
wall's equations of motion were built from a self-interacting real
scalar field with a symmetry breaking potential having a discrete set
of degenerate minima.  In our considerations we took into account two
kinds of potentials, i.e., $\phi^4$ and sine-Gordon potential.  We
used in numerical analysis simultaneously over-relaxation method
\cite{pre92} to solve equations of motion for scalar field $X$
fulfilling the certain conditions on the black hole event horizon.
The solutions of equations of motion depended on the parameter $w = 1/
\sqrt{\la} \eta $ which was responsible for the domain wall thickness
and the cosmological parameter $l$.  We paid a special attention to
extreme black hole solutions.  As in the case of RN domain wall
systems studied in Ref.~\cite{mr04} the expulsion of scalar fields
from AdS black hole is dependent on the parameter $\alpha = {r_{EH}^2
/w^2}$.  The expulsion takes place when $\alpha < 2$, while for
$\alpha > 2$ the domain wall can penetrate the extremal black hole.
We confront the numerical studies of anti-Nariai black hole domain
wall system to our very simple analytical analysis of the problem.
Contrary to simple analytical arguments given in Ref.~\cite{rog04} we
revealed that numerical studies contradicts this suggestion.  There is
no expulsion of scalar field in this case.

\begin{acknowledgments}
MR was supported in part by the Polish Ministry of Science and
Information Society Technologies grant 1 P03B 049.
\end{acknowledgments}

\bibliography{lambda}

\end{document}